\documentclass[letterpaper,twocolumn,10pt]{article}
\makeatletter
\@namedef{ver@breakurl.sty}{}%
\@namedef{ver@cite.sty}{}%
\makeatother
\usepackage[hyphens]{url} %
\usepackage{usenix}
\usepackage[sort,numbers]{natbib}
\date{}
\author{
\textnormal{Billy Bob Brumley}\\
Rochester Institute of Technology
} %

\usepackage[T1]{fontenc}
\usepackage{graphicx}
\usepackage{xspace}
\usepackage{lipsum}
\hypersetup{hypertexnames=false}
\usepackage[available,functional]{usenixbadges}
\newcommand{\PARAGRAPH}[2][.]{\smallbreak\noindent\textbf{#2#1}}

\newcommand{\CVE}[1]{\href{https://cve.mitre.org/cgi-bin/cvename.cgi?name=CVE-#1}{\mbox{CVE-#1}}}
\newcommand{\RFC}[1]{\href{https://tools.ietf.org/html/rfc#1}{RFC #1}~\citep{rfc:#1}}
\newcommand{\footurl}[1]{\footnote{\url{#1}}}

\makeatletter
\newcommand{\citeauthornolink}[1]{%
  \begingroup
    \let\hyper@linkstart\@gobbletwo%
    \let\hyper@linkend\@empty%
    \citeauthor{#1}%
  \endgroup
}
\makeatother

\title{{SoK}: The Constant Time Model}

\begin{document}

\maketitle

\begin{abstract}
Constant time programming patterns is the primary defense
against timing attacks on cryptographic implementations,
yet what ``constant time'' means varies across academia and industry.
This work systematizes constant time models and their evolution,
identifies a recurring gap between what models protect and what specifications assume,
and distills an offensive methodology for discovering timing vulnerabilities
that originate outside the cryptographic primitive boundary.
Applying this methodology,
we locate a specification-level vulnerability related to private key loading,
and confirm the leak in both OpenSSL and BoringSSL\@.
Counterintuitively,
BoringSSL's per-observation signal is several orders of magnitude stronger than OpenSSL's,
despite an explicitly stricter threat model.
 \end{abstract}

\section{Introduction}\label{sec:intro}

Timing attacks on cryptographic implementations date to the
seminal work of \citet{DBLP:conf/crypto/Kocher96},
who demonstrated that secret-dependent execution time in
modular exponentiation leaks private key material.
The three decades since have produced a rich body of work exploiting
timing leaks across increasingly sophisticated microarchitectures,
from data cache attacks on AES \cite{2005:bernstein,DBLP:conf/ctrsa/OsvikST06}
and RSA \cite{Percival05}
to speculation attacks \cite{DBLP:conf/sp/KocherHFGGHHLM019,DBLP:conf/uss/Lipp0G0HFHMKGYH18}
and beyond,
broadening timing attacks from strictly a cryptography concept
to a system security one.

The primary defense is constant time programming paradigms,
ensuring that execution flow and memory access patterns are
independent of secret data.
In practice, what ``constant time'' means varies considerably.
Multiple models have emerged over the past two decades,
each making different trade-offs between security guarantees and
practical constraints such as performance, deployability, intellectual property, and risk.
The security implications of those trade-offs are not always obvious.

An orthogonal problem is that constant time models
typically focus on cryptographic primitives in isolation, or
``crypto in a vacuum.''
Specifications that invoke these primitives
(e.g., TLS, SSH, PKCS, X.509, etc.)
routinely mandate operations on secret data
that are difficult or impossible to implement in constant time.
Historical examples include
padding validation with error handling \cite{DBLP:conf/crypto/Bleichenbacher98,DBLP:conf/eurocrypt/Vaudenay02},
variable-length encoding of secrets \cite{DBLP:conf/uss/MergetBASMS21},
and key persistence formats that leak through encoding \cite{DBLP:conf/ccs/SieckBW021}.
These specification-level vulnerabilities affect multiple implementations independently,
not due to shared code but because they all must satisfy the same requirements to achieve interoperability.

Exacerbating the situation, compilers and hardware can silently undermine
constant time programming patterns in source code.
Compiler optimizations have introduced branches on secret data
in Curve25519 \cite{DBLP:conf/cans/KaufmannPVV16}
and ML-KEM \cite{2025:arxiv:decompiling} implementations,
while variable-time instructions such as integer division
affected a wide range of Kyber
implementations \cite{DBLP:journals/tches/BernsteinBBCCKKPRT25}.
These failures exist only in compiled binaries,
invisible to source-level analysis.

\PARAGRAPH{Contributions}
This paper makes three contributions.
First, we systematize the evolution of constant time models
(\autoref{sec:models}) and compare their features, guarantees,
and documented failures (\autoref{sec:compared}),
identifying gaps between what models protect and what
specifications assume.
Second, we distill an offensive methodology
(\autoref{sec:methodology}) from a broad analysis of
specification-level vulnerability classes that provides a
systematic framework for discovering constant time violations
that originate outside the cryptographic primitive boundary.
Third, we apply this methodology to EC private key loading
(\autoref{sec:keyload}), demonstrating that both OpenSSL and
BoringSSL accept non-conformant variable-length keys and
that BoringSSL's per-observation signal
is several orders of magnitude stronger than OpenSSL's,
despite an explicitly stricter threat model.

\PARAGRAPH{Outline}
\autoref{sec:models} introduces the constant time models and traces their evolution.
\autoref{sec:compared} compares these models and identifies
gaps in empirical validation,
compiler optimization, and at the protocol level.
\autoref{sec:methodology} develops our offensive methodology
through case studies of specification-level vulnerabilities.
\autoref{sec:keyload} applies the methodology empirically to
EC private key loading.
We conclude in \autoref{sec:conclusion}.

\section{Constant Time Models: A Brief History}\label{sec:models}

The evolution of constant time models in cryptography reflects
the symbiotic relationship
between implementers focused on efficiency and attackers discovering
increasingly subtle leakage sources.
What began as simple considerations for embedded systems has evolved into a
complex landscape of threat models, each attempting to balance security
guarantees with practical requirements.
Understanding this evolution helps show why the traditional
constant time model, despite numerous attempts to relax its constraints over the years,
remains the gold standard for secure implementation.

\subsection{The Program Counter Security Model}

The earliest formal constant time considerations emerged in the context of
embedded systems and smart cards, where microarchitectures were relatively
simple and lacked complexity such as cache hierarchies and memory management units.
At this point, seminal work by \citet{DBLP:conf/crypto/Kocher96} in 1996 had already
demonstrated the risk that timing attacks pose to implementations of various public
key cryptosystems, and \citet{DBLP:conf/esmart/QuisquaterS01} in 2001 on the physical channel
side for symmetric key cryptosystems on embedded devices.

The Program Counter Security Model proposed by
\citet{DBLP:conf/icisc/MolnarPSW05} formalized the approach that was being
deployed in the embedded space at the time (2004--2005).
The model allows an attacker access to all values of the program counter as the
cryptography application executes.
If this sequence is fixed and constant, the implementation should resist timing
attacks, because constant program counter values leads to constant execution time
in their model where all instructions have fixed latency.
The authors admit this assumption is unlikely to hold outside the embedded space:
``This is not true on some architectures, due to, among other things, cache effects, data-dependent
instruction timing, and speculation'' \cite[Sec.\ 3]{DBLP:conf/icisc/MolnarPSW05}.

The rationale for allowing table lookups in this model was device driven.
In processors without caches, accessing any memory location required the same
number of cycles.
A lookup table implementation of AES S-boxes, for instance, appeared to offer
constant-time execution since each table access took identical time regardless
of the index.
A square-and-multiply-always algorithm for modular exponentiation
(RSA, Diffie-Hellman, DSA, etc.) that calls a single
multiplication function for both the square step and the multiply step would not,
in this model, allow attackers to distinguish between calls with distinct operands (multiply)
and calls with identical operands (square).
This model assumes that the primary observable characteristic was branching and
execution time at the instruction level, not the pattern of memory addresses
(containing data, not code) being accessed.

However, this simplistic model failed to account for several crucial factors,
including, for example, the observation by \citet[Sec.\ 5]{DBLP:conf/esorics/KelseySWH98}:
``We believe attacks based on cache hit ratio in large S-box ciphers \(\ldots\) are possible.''
This is not a failure in the model design per se, as a major goal of the model
was to reflect industry practice at the time.
As these implementations were ported to more complex
processors, the security assumptions of the Program Counter Security Model began to break down.

\subsection{The Traditional Constant Time Model}

The traditional constant time model emerged gradually through the work of
several researchers, rather than being defined in a single seminal paper.
The model took form
through a series of contributions that each highlighted different threats
not accounted for in the program counter model.

The 2002--2003 work by \citet{DBLP:journals/iacr/Page02} on the theoretical side and
\citet{DBLP:conf/ches/TsunooSSSM03} on the empirical side showed that
secret-dependent lookups from memory-resident tables in DES implementations
leads to key recovery attacks through cache timings.
This eventually fed into the 2004--2005 work by \citet{2005:bernstein},
a contribution on the offensive side with a data cache-timing attack on AES,
but also on the defensive side with guidelines for implementers of cryptosystems.
Bernstein warned from the data perspective:
``Using secret data as an array index is a recipe for disaster'' \cite[Sec.\ 1]{2005:bernstein}.
Even if less relevant for AES with its largely straight-line structure,
Bernstein furthermore warned from the code perspective and input-dependent branches:
``Skipping an operation is faster than doing it'' \cite[Sec.\ 9]{2005:bernstein}.
These two principles eventually became the core pillars of the traditional constant time model.

In parallel work, \citet{DBLP:conf/ctrsa/OsvikST06} introduced their
Prime+Probe and Evict+Time microarchitecture timing attack techniques,
applied to AES implementations and targeting data caches.
The authors give similar warnings as Bernstein regarding table lookups
and conditional branches, an example of the latter being with RSA
``when exponentiation is performed using a precomputed table of small powers'' \cite[Sec.\ 6]{DBLP:conf/ctrsa/OsvikST06}.
Indeed in more parallel work, \citet{Percival05} carried out such an attack on RSA,
an L1 data cache-timing attack on Intel HyperThreading architectures.
Percival concludes with similar mitigation strategies for cryptography applications:
``the code path and sequence of memory accesses are oblivious to
the data and key being used'' \cite[Sec.\ 7]{Percival05}.

In summary, the combined results and recommendations of
\citet{2005:bernstein}, \citet{DBLP:conf/ctrsa/OsvikST06}, and \citet{Percival05}
led to the traditional constant time model as we know it today.
In this light, these recommendations culminate in the 2006 design of Curve25519,
which Bernstein states
``avoids all input-dependent branches,
all input-dependent array indices,
and other instructions with input-dependent timings'' \cite[Sec.\ 1]{DBLP:conf/pkc/Bernstein06}.
Subsequent work supports the model and continues to make it explicit.
For example, the NaCl library has the design goal of
``No data flow from secrets to load addresses'' and
``No data flow from secrets to branch conditions'' \cite[Sec.\ 3]{DBLP:conf/latincrypt/BernsteinLS12}.

The traditional model thus operates under a simple but powerful principle:
the sequence of memory addresses accessed during cryptographic operations must be
identical regardless of secret values.
This includes both data addresses (no secret-dependent array indexing) and code
addresses (no secret-dependent conditional branches).
From the offensive perspective, model-wise the attacker sees all addresses
accessed while the cryptography application executes, both data and code.
As we later discuss in \autoref{sec:compared},
modern compilers have become quite skilled in recognizing related constant-time
programming patterns and instead emitting branches (see \cite{DBLP:conf/cans/KaufmannPVV16}
from 2016 and \cite[Sec.\ 2]{DBLP:journals/iacr/Pornin25a} for a wider 2025 survey),
and the model does not consider the latency of instructions being potentially operand dependent,
both of which can violate the guarantees of the traditional constant time model.

\subsection{The Intel Scatter-Gather Model}

Percival's work resulted in \CVE{2005-0109}, where Intel contributed a
patch\footurl{https://github.com/openssl/openssl/commit/ecb1445ce2df}
to OpenSSL (0.9.7h, October 2005) for hardening against timing attacks in the case of modular exponentiation.
The case is non-trivial, since high-speed implementations of Diffie-Hellman, RSA, and DSA
typically use a square-and-multiply variant of modular exponentiation with memory-resident table lookups,
costly to accomplish in constant time w.r.t.\ the traditional model.
Intel introduced a new software implementation strategy, leveraging ``scatter-gather''
to transpose data across cache lines (scatter) then assemble the data with a series of
secret-dependent table lookups (gather) that have lower than cache line granularity.
Data cache attacks at the time were only exploiting cache set access,
with no published method to determine access within a single
cache line (intra-cache-line access).

However, this approach flouts the traditional constant time model
and disregards previous advice explicitly disallowing intra-cache-line lookups.
All three of the seminal works that led to the traditional constant time model
warn against such behavior.
Percival calls it ``dangerous at best'' \cite[Sec.\ 7]{Percival05}.
Bernstein documents the related cache bank problem \cite[Sec.\ 15]{2005:bernstein},
and Osvik et al.\ worry it ``leaks information on memory accesses with resolution
better than [a cache line]'' \cite[Sec.\ 3.7]{DBLP:conf/ctrsa/OsvikST06}.

Undeterred by academic advice,
\citet{2011:ches:invited} (Intel) explicitly articulated the security model in an invited talk at CHES 2011,
stating that Intel recommends ``No secret key or data dependent:
memory access (at coarser than cache line granularity), code branching.''
Yet Intel continued to develop and push the scatter-gather technique,
for example,
2012 work by \citet{DBLP:journals/jce/Gueron12} (Intel) on efficient modular exponentiation.

To summarize the Intel model,
the fundamental assumption is that if all accesses fall within the same
sequence of cache lines, the cache behavior,
and thus the timing,
will be identical regardless of the specific location accessed within those lines.
The scatter-gather technique organizes data so that all possible accesses for a
given operation map to the same sequence of cache lines, ensuring uniform cache
behavior.
As techniques to glean intra-cache-line timings via cache bank conflicts later
emerged \cite{2013:ches:rump,DBLP:conf/ches/YaromGH16},
these days the scatter-gather technique is no longer en vogue and
the Intel model is understood to be significantly weaker than
the traditional constant time model.

\subsection{The OpenSSL Security Policy}

The OpenSSL project has developed a pragmatic threat model that focuses on the
capabilities of attackers rather than properties of the source code.
Prior to the Heartbleed vulnerability in 2014 (\CVE{2014-0160}),
OpenSSL had no explicit release strategy or security policy.
The catastrophic impact of Heartbleed,
a simple out-of-bounds read that exposed sensitive data from millions of servers,
forced a fundamental reassessment of
the project's approach to security \cite{DBLP:conf/msr/Walden20}.

Until 2019, OpenSSL would selectively issue CVEs for various types of
timing side channels (excluding physical attacks such as
power analysis \cite{DBLP:conf/crypto/KocherJJ99},
electromagnetic analysis \cite{DBLP:conf/esmart/QuisquaterS01},
fault injection \cite{DBLP:conf/crypto/BihamS97},
hardware faults \cite{DBLP:conf/isca/KimDKFLLWLM14},
data remanence \cite{DBLP:conf/uss/HaldermanSHCPCFAF08},
etc.), including those requiring colocation, such as
the CacheBleed attack \cite{DBLP:conf/ches/YaromGH16} exploiting cache bank conflicts
targeting ``constant time'' RSA in OpenSSL (\CVE{2016-0702}),
a Flush+Reload attack \cite{DBLP:conf/uss/YaromF14} on ``constant time'' P-256
arithmetic \cite{DBLP:conf/uss/GarciaB17} in OpenSSL (\CVE{2016-7056}),
or Controlled Channel attacks \cite{DBLP:conf/sp/XuCP15}
on DSA and ECDSA \cite{DBLP:conf/uss/WeiserSBS20} in OpenSSL (\CVE{2018-0734}, \CVE{2018-0735}).
However, the disclosure process was subjective and based on undocumented
aspects such as perceived risk.
This evolved further with the advent of speculative execution attacks like
Spectre \cite{DBLP:conf/sp/KocherHFGGHHLM019} and
Meltdown \cite{DBLP:conf/uss/Lipp0G0HFHMKGYH18} in 2018,
leading to the current security
policy\footurl{https://openssl-library.org/policies/general/security-policy/}
that explicitly defines which attack classes warrant CVEs and security bulletins.

The current policy attempts to strike a balance between security and practicality.
OpenSSL recognizes that all side channel threats are important, but issuing CVEs
requires significant project resources and creates downstream burdens,
particularly for FIPS-certified versions that cannot be easily modified.
Their model therefore prioritizes attacks that can be executed by less capable attackers,
i.e., those that can be mounted remotely or with limited access and resources.
The change in fact represented a major shift in the security policy of the project,
since attacks requiring colocation are not purely remote, therefore no longer captured by the model.
This excludes the vast majority of microarchitectural attacks that require local spy processes,
starting from the seminal work by \citet{Percival05}.

To summarize, OpenSSL explicitly excludes
certain attack classes from warranting security bulletins and CVEs,
including same-physical-system side channels requiring colocation,
and physical side channels such as those mentioned above.
The model assumes the offensive role and focuses on how
attackers gather data and potentially exploit vulnerabilities.

\section{Analysis: Constant Time Models Compared}\label{sec:compared}

The evolution of constant time models highlights the struggle between
theoretical security guarantees and practical implementation constraints.
Each model makes different trade-offs, and history has shown that deviations
from the most conservative approach consistently introduce vulnerabilities.
\autoref{tab:compared} systematizes these
features, guarantees, constraints, trade-offs, deviations, and risks.
This section consists of a discussion centered around the contents of \autoref{tab:compared}.

\begin{table*}
\caption{Constant time models compared across rules, scope, validation, and documented failures.
Every relaxation of the traditional model has eventually been exploited.}%
\label{tab:compared}
\phantom{z}
\centering
\resizebox{1.0\linewidth}{!}{%
\begin{tabular}{|l|l|%
p{0.32\linewidth}|%
p{0.32\linewidth}|%
l|%
p{0.18\linewidth}|%
p{0.2\linewidth}|} \hline%
\textbf{Model} &
\textbf{Year} &
\textbf{Memory Access Rules} &
\textbf{Control Flow Rules} &
\textbf{Protocol-Layer Scope} &
\textbf{Validation Approach} &
\textbf{Why it Failed} \\ \hline%
Program Counter &
2004--2005 &
Allows data flow from secrets to load addresses &
No data flow from secrets to branch conditions &
Crypto primitives only &
Differential analysis of PC sequences &
Data cache timing attacks \\ \hline%
Traditional CT &
2004--2006 &
No data flow from secrets to load addresses &
No data flow from secrets to branch conditions &
Crypto primitives only &
Static code analysis &
Compiler optimizations, variable-time instructions \\ \hline%
Intel Scatter-Gather &
2005 &
Allows data flow from secrets to load addresses (within cache lines) &
No data flow from secrets to branch conditions &
Crypto primitives only &
Microarchitectural assumptions &
Cache bank conflicts (CacheBleed) \\ \hline%
OpenSSL &
2019 &
Context-dependent restrictions &
Context-dependent restrictions &
Protocol aware &
Threat modeling, empirical data &
Limited threat model scope \\ \hline%
BoringSSL &
2025 &
Allows data flow from secrets to load addresses (variable-length inputs) &
Allows data flow from secrets to branch conditions (variable-length inputs) &
Crypto primitives only &
Static code analysis, formal methods &
This work \\ \hline%
\end{tabular}
}
\end{table*}

\subsection{Model-Specific Analysis}

\PARAGRAPH{Program Counter}
This model assumes that instruction timing is predictable and data-independent.
While largely suitable for simple embedded systems without caches, the model
immediately fails when applied to large classes of modern processors.
Three seminal attacks invalidated this model:
(i) the AES data cache attack by \citet{DBLP:conf/ctrsa/OsvikST06},
demonstrating that data cache attacks could reveal the complete granular
memory access pattern to S-boxes through timing analysis;
(ii) the RSA data cache attack by \citet{Percival05},
a practical attack on HyperThreading architectures leading to complete
private key recovery via highly granular L1 data cache timing data;
(iii) the AES data cache attack by \citet{2005:bernstein},
a purely statistical attack focusing on threat models involving remote attackers
with only coarse timing data.
The fundamental flaw in the model was defining table lookups with secret indices as safe,
an assumption that fails when applied to the vast majority of COTS
processors today that typically feature several layers of cache memory.
The model's validation approach of analyzing PC sequences
indeed allows verification that no secret-dependent branches occur,
yet fails to capture the threat of secret-dependent data access.

\PARAGRAPH{Traditional}
The main focus of the traditional constant time model is on the source code itself.
The model considers only a single seemingly simple requirement:
all addresses are public,
the consequence being that all code must
ensure no branching on secrets and no data access using secret indices.
These features are
typically accomplished with a combination of side channel friendly algorithm
selection and (now) well-understood constant time programming paradigms.
Regardless of its theoretical soundness, the traditional model faces many practical challenges.

First, compiler optimizations can silently counteract constant time programming patterns.
Work by \citet{DBLP:conf/cans/KaufmannPVV16} demonstrated how constant time source code
for Curve25519 became variable time after compilation.
These implementations contain several multiplications with public curve-related constants,
and Microsoft's MSVC 2015 toolchain expanded one such multiplication by a 32-bit constant
inside a subroutine using a branch on the sign of the second operand, since the upper
32 bits of said constant are known to be zero a priori.
The authors outline a method for complete private key recovery leveraging this leak,
using modest computation power.

Second, certain instructions have input-dependent timing.
An excellent example is early-termination multipliers on ARM processors \cite{DBLP:conf/icisc/GrossschadlOPT09}.
Multiplication unit designs in older 32-bit ARM chips leveraged smaller
parallel multipliers that would exit early when encountering leading zero
bytes in one of the operands.
This meant the latency of a multiplication instruction would vary between 2--5 clock cycles,
depending on the number of leading zero bytes.
A more recent example is the KyberSlash attack \cite{DBLP:journals/tches/BernsteinBBCCKKPRT25}.
Many Kyber implementations feature an integer division by a small public constant,
and expect the compiler to either replace that with a multiplication by a constant or
a small number of integer shifts and integer additions.
Yet some compilers with appropriate optimization flags would faithfully emit
an integer division instruction, which is notoriously variable time on Intel architectures.
KyberSlash was both timely and impactful, discovered around the time of NIST standardization
and applying to a large portion of software implementations,
including the reference implementation by the Kyber authors themselves.
It continues to resurface, e.g., \CVE{2026-22705} in RustCrypto's ML-DSA implementation.

In summary, despite its age,
the traditional constant time model has stood the test of time and
proves to be the de facto standard for secure implementations.

\PARAGRAPH{Intel Scatter-Gather}
Intel's constant time model attempts to balance security and performance.
It is conceptually similar to the traditional constant time model,
yet allows data flow from secrets to addresses at cache line granularity \cite{2011:ches:invited}.
Put another way, in the traditional model where all addresses are public,
in the Intel model all addresses sans the least significant \(\log_2(\delta)\) bits
are public (where \(\delta\) is the number of bytes in a cache line).
This relaxation was based on the assumption that accesses within the same cache
line would have fixed timing.
At least three works demonstrate concrete failure of the Intel model.

First, \citet{2013:ches:rump} presented empirical data that showed
data-dependent timing variations by artificially inducing cache bank conflicts.
While the experiment in question targeted strawman code and not actual cryptography code,
the fundamental microarchitectural effect was clear from the empirical data.

The CacheBleed attack built on that proof-of-concept
soon thereafter \cite{DBLP:conf/ches/YaromGH16}.
The code targeted by CacheBleed was in fact a contribution from
Intel as a countermeasure in OpenSSL to the attack by \citet{Percival05}.
This scatter-gather code used secret-dependent table lookups
within cache lines during modular exponentiation,
which the CacheBleed attack exploited due to
timing variations caused by cache bank conflicts.
The authors were able to recover an RSA private key
using a modest number of signature queries.

Lastly, even secret-dependent control flow within a cache line
is not a secure strategy.
The PortSmash attack (\CVE{2018-5407}) exploited port contention in
Simultaneous Multithreading (SMT) architectures
(such as Intel's HyperThreading)
to leak control flow \cite{DBLP:conf/sp/AldayaBHGT19}.
One of the experiments the authors conducted targets code with
multiple branches within a single cache line,
and PortSmash is still able to distinguish between the
branch targets \cite[Sec.\ 4]{DBLP:conf/sp/AldayaBHGT19}.

In summary, cache bank conflicts prophetically invalidated
Intel's constant time model for data, and port contention
accomplished the same for code.
The Intel model nevertheless serves as an exemplary case study of why
deviating from the traditional constant time model carries significant risk.

\PARAGRAPH{OpenSSL}
OpenSSL's pragmatic, data-driven approach focuses on attacker capabilities
and empirical data rather than source code analysis.
Hardware-based side channels and any attack requiring co-location
(e.g., most microarchitecture attacks) are out of scope.
With these restrictions, OpenSSL is hoping to exclude local attackers
from their model, yet retain remote attackers.
This model fails in predictable ways
(e.g., the vast majority of cache attacks),
yet also fails in surprising ways, discussed below.

\citet{DBLP:conf/uss/WeiserSBS20} discovered several vulnerabilities in
multiprecision integer arithmetic libraries used by cryptography libraries.
Their ``V8'' vulnerability applied to a division function as part of the
extended Euclidean algorithm, their main application being to DSA signing.
The leak was due to data structure resizing, and was limited to a
single top-most bit of the secret.
Their ``V9'' vulnerability is in the same algorithm,
a conditional negation that leaks the top-most bit of a secret inverse.
These vulnerabilities are outside the scope of the OpenSSL security policy because
(i) they are unlikely to be observable remotely; and
(ii) it is unclear from the theory perspective how a remote attacker
can effectively utilize such a small leak.
Hence, while the vulnerabilities are certainly valid from the academic perspective,
OpenSSL's security policy discards them.

A second example is the obscure key formats studied by \citet{DBLP:conf/uss/GarciaHTGAB20}.
The code loading these keys neglected to configure the private keys to ensure
constant time code paths within the library, which led to a single trace private key
recovery attack with local L1 data cache timings.
While that is clearly outside OpenSSL's model, the same leak revealed information
about the top set bit of the private key, potentially measurable in remote scenarios.
Yet it is again unclear from the theory perspective how a remote attacker
can effectively utilize such a small leak on fixed bits.
Hence, while this vulnerability is certainly valid from the academic perspective,
it is out of scope for OpenSSL's security policy.

A third example is the LadderLeak attack by \citet{DBLP:conf/ccs/AranhaN0TY20},
capable of recovering ECDSA private keys with less than one bit of nonce
leakage per signature, but requiring local cache timings to detect branch targets.
Coupled with the large data requirement of the attack
(needing approximately \(2^{33}\) signatures for key recovery),
LadderLeak falls outside OpenSSL's security policy,
requiring local access and significant data, time, and resources.
Yet again, this attack is certainly valid from the academic perspective.

A final and recent example is the SLasH-DSA attack by \citet{journals/uasc/BoyPPWE26},
a fault attack on hash-based PQC as implemented in OpenSSL 3.5.
The authors utilize the RowHammer technique \cite{DBLP:conf/isca/KimDKFLLWLM14} to induce
memory read faults while querying digital signatures.
Hardware faults are indeed explicitly excluded in OpenSSL's security policy,
and furthermore RowHammer requires local access.
Yet again, this attack is certainly valid from the academic perspective.

An interesting model-related case study is
the OpenSSL vulnerability linked to \CVE{2025-27587},
which led to conflict between OpenSSL and the wider security community.
The vulnerability leaks part of the ECDSA
nonce in OpenSSL's implementation on PowerPC architectures,
yet the timing distributions are highly overlapping and
it remains unclear how to separate them.
Furthermore, the experiment data is with local timings,
while remote timings exacerbate the situation from the attacker perspective,
increasing the statistical overlap.
OpenSSL typically manages its own CVEs,
but in this case MITRE chose to issue its own against OpenSSL,
who disputed the vulnerability because it fell outside their written security policy,
for similar reasons as the LadderLeak attack.

In summary, OpenSSL's model has the best and worst of both worlds.
On one hand, it flippantly excludes large classes of attacks that the security
community historically recognizes as valid, i.e., requiring co-location and local execution.
On the other hand, their data-driven approach potentially scopes in
threats that are typically not captured in other models,
where ``crypto in a vacuum'' is the norm,
source code analysis stops at the low-level crypto,
and vulnerabilities in callers to the crypto are another layer's problem,
even if attackers care little about such semantics.
We elaborate in the next subsection.

\PARAGRAPH{BoringSSL}
We defer this discussion to the end of the paper.

\subsection{Constant Time Models: Gaps}

Following the previous analysis surrounding \autoref{tab:compared},
in this subsection we expand the discussion to consider areas
where the models may not perform as expected.
These issues are pervasive,
and represent a security concern for all of the models.

\PARAGRAPH{Protocol-layer leakage}
A critical limitation of most constant time models is their focus on
cryptographic primitives and low-level arithmetic in isolation,
ignoring how these primitives behave when integrated into protocols.
This gap has led to numerous attacks that bypass even high-assurance and
high-security cryptography implementations by exploiting protocol-level leaks.
While we expand later in \autoref{sec:methodology},
here are a few seminal works that highlight this gap.

The Million Message Attack (MMA) by \citet{DBLP:conf/crypto/Bleichenbacher98}
demonstrated that implementing RSA PKCS\#1~v1.5 (\RFC{2313}, 1998)
carries significant security risk.
Implementations tend to leak whether decryptions produce valid padding or not,
either directly through error messages or indirectly through timing.
These leaks enable adaptive chosen ciphertext attacks to decrypt without the private key.
\citet{DBLP:conf/crypto/Manger01} subsequently showed that RSA-OAEP and
PKCS\#1~v2.0 (\RFC{2437}, 1998),
designed to resist MMA,
remained vulnerable to similar oracle attacks through timing leaks
in their more complex padding validation.

On the symmetric side, \citet{DBLP:conf/eurocrypt/Vaudenay02} targeted
implementations of CBC mode for block ciphers with PKCS\#7 padding,
where timing differences in padding validation expose the plaintext.
While \citeauthornolink{DBLP:conf/eurocrypt/Vaudenay02} described several different
oracles, the general area of padding oracles emerged only later,
encompassing both the attacks by \citeauthornolink{DBLP:conf/crypto/Bleichenbacher98} and
\citeauthornolink{DBLP:conf/eurocrypt/Vaudenay02} under the general umbrella that
validating padding applied to plaintext is a dubious security practice.
The work extends naturally to hardware implementations \cite{DBLP:conf/crypto/BardouFKSST12}.

These seminal results reveal a fundamental class of vulnerabilities: protocols
requiring conditional processing of padding or error conditions create timing
channels that propagate beyond the cryptographic primitive layer.

\PARAGRAPH{Empirical validation}
Consulting the rightmost column of \autoref{tab:compared},
the failures across all models highlight the critical importance of empirical validation,
both through direct measurement of execution latency on target hardware and
through analysis of compiled binaries.
Even when source code follows constant-time programming practices,
the underlying hardware may still exhibit input-dependent timing variations,
and mainstream compilers actively undermine constant time guarantees.

The tool dudect \cite{DBLP:conf/date/ReparazBV17} takes a purely empirical approach,
which utilizes classical leakage detection methods.
It executes the target function with two classes of input data,
collects cycle-accurate timings,
and applies statistical tests to determine whether the
resulting distributions are distinguishable.
Because this approach requires no hardware model and no source code annotations,
any timing variation observable on the target platform can in principle be detected,
provided the effect is large enough relative to measurement noise.
However, the tool cannot pinpoint the source of a detected leak,
and subtle violations may require millions of measurements to reach statistical significance.

Constant time violations can originate in the toolchain rather than the hardware.
The Clangover vulnerability (\CVE{2024-37880}, Antoon Purnal) in the ML-KEM reference
implementation demonstrates a case where the source code used a
constant-time conditional assignment, but modern Clang compiled it into a conditional
branch on secret data~\cite{2025:arxiv:decompiling}.
The vulnerability is particularly stealthy because source-level analysis tools
correctly report the code as constant-time,
since the violation exists only in the compiled binary.
LLVM recently implemented support for source code annotations that
preserve select constant-time properties through optimization
passes\footurl{https://blog.trailofbits.com/2025/12/02/introducing-constant-time-support-for-llvm-to-protect-cryptographic-code/},
demonstrating that compiler designers are aware of the issue,
yet the narrow scope and opt-in nature make this approach inherently fragile.
\citet{2025:arxiv:decompiling} further showed that popular decompilers compound the
problem by performing optimizations that eliminate constant time violations during
analysis, causing decompile-then-analyze approaches to miss vulnerabilities that
exist in deployed binaries.
In summary,
this is a challenging compiler security gap since developers write
correct constant-time code, but the toolchain silently transforms it into
vulnerable binaries, and standard analysis techniques fail to detect the
resulting violations.

To approach this challenge,
TIMECOP uses Valgrind's Memcheck infrastructure to detect potential
constant-time violations through dynamic binary analysis.
It marks secret data as uninitialized via Valgrind client requests,
causing Memcheck to flag any secret-dependent branch conditions or
memory accesses encountered during execution.
Because this analysis operates on compiled binaries,
it can detect violations introduced by compiler optimizations that
are not visible at the source level.
TIMECOP originated as prototype patches to the SUPERCOP benchmarking framework,
with a leakage model limited to branching and memory accesses.
From the other direction, patches to Valgrind added support for detecting
potential variable-time instructions operating on secret data.
Finally, the KyberSlash work \cite{DBLP:journals/tches/BernsteinBBCCKKPRT25}
combines both approaches to allow the analysis to scale by
scoping in more ISAs,
extending the list of potentially vulnerable instructions, and
enabling tests across a wide spectrum of implementations and architectures.

In summary,
the two approaches complement one another,
with dudect analyzing end-to-end timing behavior on real hardware,
while binary analysis tools like TIMECOP pinpoint the location of the leak.
\section{An Offensive Security Methodology}\label{sec:methodology}

Most of the constant time models in \autoref{sec:compared} stop at the crypto layer,
yet the protocols using these primitives often operate on secret data well beyond that boundary.
Specifications, standards, and protocols are typically written with correctness and
interoperability as primary goals, not constant time execution.
They often require operations that create natural timing channels, such as
padding schemes with conditional processing,
error handling with different code paths, and
variable-length encodings.
No implementation can satisfy both the specification and constant time requirements,
because the specification itself mandates contradicting behavior.

This section explores those gaps through various case studies,
culminating in an offensive methodology which
identifies standards whose flexibility or complexity creates said conflict,
then tests implementations for any leaks predicted by the analysis.
We organize the case studies by the specification at fault,
not by the models from \autoref{tab:compared}.
Which constant time model a given attack violates depends on the threat model,
while the specification does not.
Each case below is an instance of protocol-layer leakage.

\subsection{Case Studies}

Specifications have been a steady source of timing leaks since (at least) the
seminal attacks discussed in \autoref{sec:compared},
with new instances appearing across protocols and implementations in the decades since.
Standards and specifications play a significant role in proliferating these vulnerabilities,
and in this section we investigate that relationship
through several recurring cases
(e.g., error handling, padding validation, key persistence formats, encoding of secrets)
where compliant implementations undermine primitive-level constant time guarantees with
system-level violations.

\PARAGRAPH{Public key padding oracles}
As previously discussed,
the way that PKCS\#1~v1.5 (\RFC{2313}) specifies the handling of malformed
ciphertexts lends itself to vulnerability,
first exploited by \citet{DBLP:conf/crypto/Bleichenbacher98} in 1998.
Each fix has produced a new variant,
several of those which we now discuss.

\citet{DBLP:conf/esorics/JagerSS12} observed that the XML encryption standard
specified its own error handling requirements for RSA PKCS\#1~v1.5,
independently recreating the same class of oracle in a completely
different protocol (\CVE{2011-2487}).
This is despite the W3C having published the specification in 2002,
four years after Bleichenbacher's original disclosure.

\citet{DBLP:conf/woot/KupserMSS15} subsequently automated the attack,
finding that several implementations (\CVE{2015-0226}, etc.)
remained vulnerable even after deploying mitigations.
The authors circumvent fixes via wrapping techniques that relocate encrypted
elements within the XML document tree to bypass integrity checks.

The ROBOT attack \cite{DBLP:conf/uss/BockSY18} (\CVE{2017-6168}, etc.) demonstrated that even the
TLS~1.2 specification's own countermeasure against Bleichenbacher
was itself a source of new oracles.
The mitigation requires servers to generate a random premaster secret on
padding failure and continue the handshake.
The work highlights that ensuring the success and failure codepaths are
indistinguishable is indeed difficult to achieve in constant time.

\citet{DBLP:conf/sp/RonenGGSWY19} identified nine distinct
attack variations across the TLS ecosystem,
each exploiting different leakage sources in implementations with mitigations
(\CVE{2018-12404}, etc.).
Each fix to a specific Bleichenbacher variant introduced new,
subtly different oracles exploited by the authors.
This shows that when a specification is fundamentally incompatible with constant time goals,
patches merely shift the leak.

\PARAGRAPH{Symmetric key padding oracles}
The 2002 padding oracle attack by \citet{DBLP:conf/eurocrypt/Vaudenay02},
mentioned in \autoref{sec:compared},
is the symmetric key counterpart.
The specification-level root cause
(protocols that require the receiver to act on plaintext properties before integrity can be verified)
has proven equally persistent,
with instances across SSH, XML encryption, and TLS\null.

\citet{DBLP:conf/sp/AlbrechtPW09} analyzed the SSH Binary Packet Protocol (BPP, \RFC{4253}),
finding that it encrypts the packet length field within the first CBC ciphertext block
(\CVE{2008-5161}, etc.).
The receiver must decrypt this block to see how many additional bytes to read.
The specification makes it impossible to authenticate the MAC before parsing this field,
a parallel to the (legacy) TLS MAC-then-encrypt ordering.
Observing how many bytes the server reads before responding
allows an attacker to infer the decrypted length field,
yielding a practical chosen-ciphertext attack enabled through the standard.

\citet{DBLP:conf/ccs/JagerJ11} showed that the XML encryption
specification's use of CBC mode created a similar oracle,
exploiting the server's error handling when decrypted
plaintext contained malformed character encodings.
This vulnerability (\CVE{2011-1096}) originated from the XML parsing layer
rather than cryptographic padding validation.

Previously mentioned in the public key context,
\citet{DBLP:conf/esorics/JagerSS12} further showed that
XML encryption combines the two oracles on the symmetric key and public key sides.
CBC malleability allowed an attacker to test candidate RSA session keys by
observing whether the resulting symmetric decryption produced valid plaintext,
enabling Bleichenbacher attacks without relying on
error messages or timing from the RSA layer itself.

The Lucky Thirteen attack \cite{DBLP:conf/sp/AlFardanP13} (\CVE{2013-0169}) extended
the CBC work to (pre 1.3) TLS with a timing attack on the MAC computation itself.
The number of hash iterations depends on the data length after stripping padding (mandated by the specification),
leaking whether a modified ciphertext contained valid padding through timing.
The attack affected virtually every TLS implementation because the leak
originated in the specification's ordering of operations, not in any
particular cryptographic primitive.
\citet{DBLP:conf/ccs/ApececheaIES15} subsequently showed that cache-based
timing attacks could amplify these differences with local access.

\PARAGRAPH{Encoding leakage}
Beyond padding validation and error handling, specifications that define
key persistence formats are another source of timing leaks.
Encodings such as hex and base64
(required by any library that loads cryptographic material from PEM files or similar formats)
typically use a lookup table indexed by the ASCII value of each input character,
easily leading to secret-dependent memory access,
and furthermore entirely outside the constant time cryptographic primitive boundary.

\citet{DBLP:conf/ccs/SieckBW021} demonstrated this concretely,
showing that base64 decoding in OpenSSL, mbedTLS (\CVE{2021-24119}),
Botan (\CVE{2021-24115}), Mozilla NSS, and GNU Nettle used secret-dependent memory accesses
when loading RSA private keys from PEM files.
Observing a single PEM key loading event allowed partial recovery of
RSA private key material,
leakage that no amount of constant time arithmetic in the
cryptographic primitive layer can mitigate.
Subsequent work increased the leakage with finer spatial resolution
through advanced microarchitectural techniques \cite{DBLP:journals/tches/SieckZBCEY24}.

Notably, BoringSSL had proactively addressed this leak
in 2017 before either attack was published,
implementing constant time base64 decoding by replacing table lookups with
arithmetic range comparisons.
Quoting the commit message:\footnote{\url{https://boringssl.googlesource.com/boringssl/+/536036ab}}
``PEM files sometimes carry private keys so, in principle,
we'd probably prefer not to leak the contents when we encode or decode them.''
BoringSSL chose to extend constant time techniques to key decoding
well before any published attack,
acknowledging that decoding can leak secret material and applying meaningful
mitigations beyond the cryptographic primitive boundary.

\PARAGRAPH{Canonicalization leakage}
Beyond character-level encoding,
some specifications also mandate canonical representations for secret values.
This might include
stripping leading zeros,
conditionally prepending sign bytes (e.g., for DER),
or enforcing a particular integer representation.
Compliant implementations would then perform variable-length processing on secret data,
potentially leaking the length of the secret
(confidential in many cases).

A good example is the SSH protocol (\RFC{4253}, discussed earlier in the separate context of BPP),
where the standard mandates that the Diffie-Hellman shared secret be represented as an MPINT\null.
This data type is defined in \RFC{4251} with the explicit requirement that
``unnecessary leading bytes with the value 0 \ldots{} MUST NOT be included.''
\RFC{8731} (which defines Curve25519 and Curve448 key exchange for SSH)
explicitly acknowledges this weakness in its Security Considerations section:
``The way the derived MPINT binary secret string is encoded before it
is hashed (i.e., adding or removing zero bytes for encoding) raises
the potential for a side-channel attack.''

The Raccoon attack by \citet{DBLP:conf/uss/MergetBASMS21} exploited similar
canonicalization in (pre 1.3) TLS,
where DH\null(E) cipher suites require stripping leading zeros from the
premaster secret before key derivation.
The attack affected several security-critical libraries
(\CVE{2020-1968} in OpenSSL, \CVE{2020-12413} in Mozilla NSS, etc.),
not due to shared code, but because they implemented the same specification.
Subsequent work~\cite{DBLP:conf/uss/AldayaB22} demonstrated that
microarchitectural timings from local attackers could amplify the leakage,
extending it to CMS and S/MIME contexts.
The Raccoon authors confirmed that SSH suffers from the same specification-level flaw (discussed previously),
but could not exploit it due to SSH's strict use of fresh keys per session \cite[Sec.\ 8]{DBLP:conf/uss/MergetBASMS21}.
Exploitability depends on the protocol context,
even when the underlying specification-level vulnerability is identical.

\subsection{The Offensive Methodology}

In each of the previous case studies,
the specification requires an operation on secrets
that is inherently difficult to realize in constant time.
Implementations that comply for interoperability leak in
ways that no amount of constant time arithmetic or programming
paradigms in the cryptographic primitive layer can prevent.
We now distill this into a methodology.

\PARAGRAPH{Step 1: Identify specs with problematic operations}
Locate operations a specification requires on secret data
that conflict with constant time goals.
For example,
padding schemes with conditional processing (PKCS\#1 v1.5),
error conditions that mandate distinguishable responses (CBC),
variable-length encodings (ASN.1, DER, PEM),
canonicalization of keys (byte stripping),
etc.

\PARAGRAPH{Step 2: Analyze implementation strategies}
Investigate how systems handle these problematic operations.
For example,
they might knowingly tolerate the leak,
deploy countermeasures that might expose corner cases,
or compromise on compatibility to achieve better timing attack resistance.
Implementation strategies may also vary across architectures
and compiler toolchains,
so the same code can exhibit different leakage characteristics
depending on the target platform.

\PARAGRAPH{Step 3: Confirm the analysis}
Run experiments against these implementations
to confirm or refute the predicted leak.
The systematization in \autoref{sec:compared}
suggests that multiple analysis techniques apply in tandem.
The scope, validation, and failure considerations in \autoref{tab:compared}
highlight
static methods,
dynamic analysis,
and targeted timing measurements
as examples of analytical methods.
Casting a wide net allows leveraging a fuller spectrum of evidence,
from more formal techniques all the way through empirical timings.

In summary,
different implementations of the same specification tend to leak in similar ways,
not because they share code,
but because they share the specification.
Diverse analysis methods complement one another,
and our methodology is sound largely due to inheriting more of the
strengths of these methods and less of the inherent limitations.
We validate our methodology in the remainder of this paper.
\section{Analysis: Private Key Loading}\label{sec:keyload}

Applying said methodology,
this case study examines how a seemingly minor specification detail,
the encoding of elliptic curve private keys,
creates systematic timing leaks across implementations.
By tracing the evolution from early standards through modern implementations,
we demonstrate how specification-level decisions made decades ago continue to
create meaningful timing leaks today.

\subsection{Elliptic Curve Cryptography Standards}
The encoding of elliptic curve private keys dates to the
Standards for Efficient Cryptography Group's specification
SEC~1 \cite{sec1v1}, published in 2000, which defined the
\texttt{ECPrivateKey} syntax with the private key encoded as a
fixed-length octet string derived from the Integer-to-Octet-String-Primitive (I2OSP).
\RFC{5915} later echoed this requirement,
specifying that the \texttt{privateKey} field must be of length
\( \lceil \log_2(n) / 8 \rceil \) bytes, where \( n \) is a public constant (the curve order).
For example, with the NIST standard curves P-256, P-384, and P-521,
this mandates exactly 32 bytes, 48 bytes, and 66 bytes, respectively,
regardless of the numeric value of the scalar,
and any private key with a small value must be zero-padded to the full length.
This fixed-length requirement happens to enable constant-time processing,
as decoders can unconditionally read a publicly known number of bytes and convert
them to an internal BIGNUM representation using fixed-width operations.

In practice, implementations are not always so rigid.
The \texttt{privateKey} field is a DER-encoded OCTET STRING,
but the value it carries is semantically an integer,
and internally many implementations will convert it to a BIGNUM\@.
DER encoding of INTEGER values requires minimal-length representation,
and per X.690 rules,
the output must use the smallest number of octets needed,
canonicalization that strips leading zero bytes.
While OCTET STRING and INTEGER are distinct ASN.1 types with different
encoding rules, the boundary might blur in practice.
BIGNUM libraries often strip leading zeros on import regardless of the source type,
in which case subsequent serialization routines naturally produce minimal-length output.
Ecosystem-wise, the result would be a population of existing keys with
private key fields shorter than the specification mandates,
encoding the effective (bit, byte) length of the private key (secret integer scalar)
directly in the serialized representation.
Implementations could choose to reject these non-conformant keys,
yet lessons from padding oracle attacks tell us this
is more difficult to do securely than it might seem.

Applying the \autoref{sec:methodology} methodology,
the first step is to confirm that the target
implementations accept non-conformant keys at all.
We generated P-384 private keys with effective bit lengths ranging from
128 to 384 in multiples of 8,
encoding each as a truncated (leading-zeros-stripped) octet string within
an otherwise valid \texttt{ECPrivateKey} structure.
Both OpenSSL and BoringSSL accept every such key without error,
confirming that the threat surface exists and justifying deeper analysis
of the loading paths.

In summary, this suggests canonicalization leakage
(previously discussed in \autoref{sec:methodology}),
where a specification defines a fixed-length format,
but implementations that tolerate variable-length inputs potentially introduce
vulnerabilities outside the cryptographic primitive boundary.

\subsection{Implementation Analysis}
Having confirmed that both implementations accept variable-length keys,
we now examine their respective loading paths to understand where
secret-dependent processing occurs.
We also discuss adversarial scenarios and
the high level impact to security.

\PARAGRAPH{OpenSSL}
EC private key loading follows the path
\texttt{d2i\_ECPrivateKey} \( \to \)
\texttt{EC\_KEY\_oct2priv} \( \to \)
\texttt{ossl\_ec\_key\_simple\_oct2priv}.
Two operations process the variable-length private key material.
First, OpenSSL's ASN.1 runtime (\texttt{d2i\_EC\_PRIVATEKEY}) decodes the DER structure,
populating an \texttt{ASN1\_OCTET\_STRING} whose length field
reflects the encoded size of the private key.
For a non-conformant key this is shorter than the curve order,
and the parser performs correspondingly less work.
Second, \texttt{ossl\_ec\_key\_simple\_oct2priv} passes that
variable-length buffer and its length directly to
\texttt{BN\_bin2bn} to construct a BIGNUM holding the private key,
which is not a constant-time function.
It processes exactly the number of bytes present,
so the amount of work varies with the encoded length of the private key.

\PARAGRAPH{BoringSSL}
BoringSSL's loading path is architecturally similar.
The function \texttt{ec\_key\_parse\_private\_key} handles both ASN.1
parsing and key construction in a single routine,
using BoringSSL's CBS (``crypto byte string'') interface for DER decoding.
As with OpenSSL, two operations process the variable-length private key
material.
First, \texttt{CBS\_get\_asn1} extracts the OCTET STRING from the DER
structure, parsing a TLV whose length field encodes the size of the
private key material.
For a non-conformant key, this length is shorter than the curve order,
and the parser performs correspondingly less work.
Second, the extracted bytes are passed to \texttt{BN\_bin2bn}
 to construct a BIGNUM holding the private key,
which as in OpenSSL processes exactly the number of bytes present
with no padding to a fixed width.
Both leakage points (the ASN.1 length parsing and the BIGNUM conversion)
operate on the actual encoded length rather than the expected (public) curve order length.

\PARAGRAPH{Security implications}
Some services amortize the cost of key loading using
daemons that load keys once at startup and serve from memory.
Others load per connection,
a good example being the classical Unix service dispatcher
inetd (``internet service daemon'') where (typically)
each incoming connection spawns a fresh process
that parses its keys before serving the request.
In the latter case,
an adversary can force an arbitrary number of
key load observations simply by opening connections.

For this leak,
increasing samples does not provide new information in an unbounded way.
The leak reveals fixed information,
the byte length of the encoded scalar,
which is a property of the long-term private key,
not the session.
Additional queries only increase confidence.
To give a worked example,
probabilistically \(1/256\) keys will be a single byte shorter than a full-length key,
and in that case it leaks 8 bits (``the top byte is zero'').
So while this leak indeed leads to an information-theoretic reduction in confidentiality,
it cannot be used to extract
a growing fraction of the fixed secret.

The utility of the leak from the adversarial perspective is also dubious.
Continuing the worked example,
any cryptanalytic attack that benefits from a short fixed scalar can
simply assume one and proceed,
without any leak or side channel.
Such an attack succeeds \(1/256\) of the time.
While the leak is real and any model violations are real,
we are not aware of any exploit,
and see little to no practical relevance in the leak.

\subsection{Results}\label{sec:keyload_results}
We constructed a harness that loads each PEM-encoded key file into memory once,
then repeatedly parses it from an in-memory buffer.
The timed region brackets a single call to
\texttt{PEM\_read\_bio\_PrivateKey}
(PEM base64 decoding, DER parsing, and key object construction),
with no disk I/O in the measurement path.
The harness runs on a quad-core Intel i5-7500T at 2.70GHz with frequency scaling disabled,
equipped with 16GB of DDR and running Debian 12.13 ``bookworm.''
This CPU does not feature HyperThreading.
It loads each key 1,000 times, each load timed in CPU clock cycles.
We linked the same harness against both OpenSSL and BoringSSL,
using a corpus of P-384 keys at 33 distinct bit lengths
(128 to 384, multiples of 8) with equal sample counts per class.
This uniform distribution is artificial
(a randomly generated P-384 scalar has probability roughly 255/256
of occupying the full 48 bytes),
but allows
characterizing the timing distribution by class.

\begin{figure*}[!ht]
\centering
\includegraphics[width=0.33\linewidth]{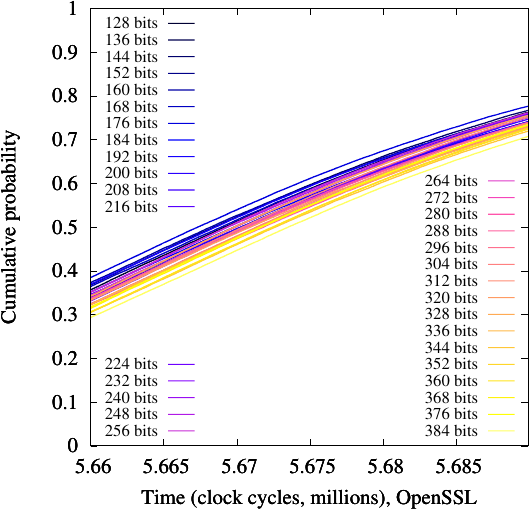}\hfill%
\includegraphics[width=0.33\linewidth]{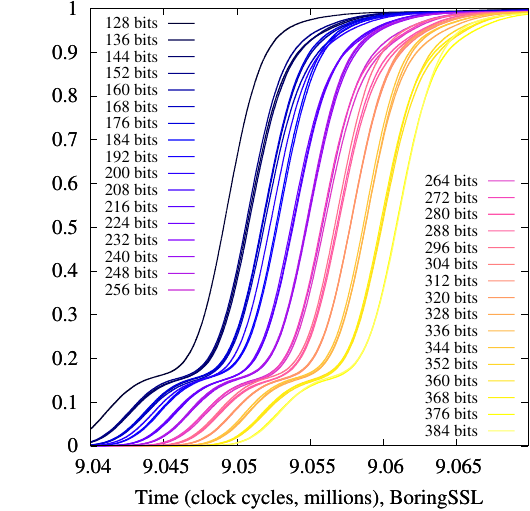}\hfill%
\includegraphics[width=0.33\linewidth]{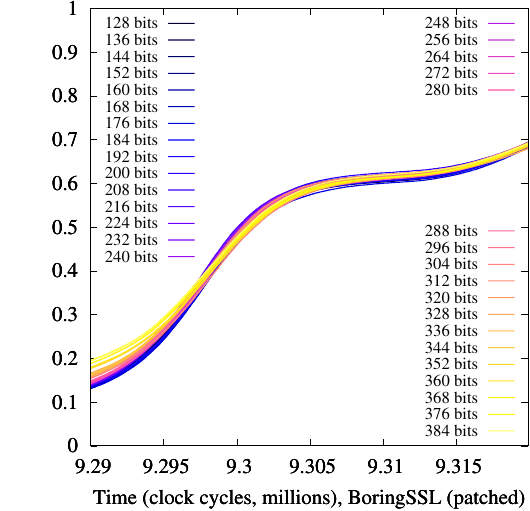}
\caption{Empirical CDFs of EC private key loading time (P-384)
grouped by effective byte length.
OpenSSL (left): distributions overlap almost entirely.
BoringSSL unpatched (center): distributions separate visibly,
with a staircase structure at byte granularity.
BoringSSL patched (right): with the hedge, distributions overlap,
similar to OpenSSL.}%
\label{fig:keydist_boringssl}
\end{figure*}

\autoref{fig:keydist_boringssl} shows the empirical CDFs of
loading time grouped by encoded byte length.
The OpenSSL distributions (left) overlap almost entirely,
with no visually discernible separation.
The BoringSSL distributions (center) separate visibly,
with a staircase structure suggesting leakage at byte granularity.
\autoref{tab:keydist_stats} summarizes the statistical comparison
across three complementary metrics.
The mutual information \( I(\mathrm{bit length};\mathrm{timing})\)
across all 33 classes measures how much key-length information
a single timing observation reveals.
OpenSSL leaks 0.0031 bits of key-length information per timing
observation; BoringSSL leaks 0.6104 bits, approximately 200\(\times\) more.
To test the null hypothesis and establish a noise baseline,
we shuffled class labels and
recomputed MI over 100 rounds;
the observed MI for all configurations exceeds this baseline,
confirming the signal is genuine,
though the magnitude differs significantly.

\begin{table}[t]
\caption{Statistical comparison of key load timing leakage
for OpenSSL, BoringSSL (unpatched and hedged), NIST curve P-384.
MI and permutation \(z\)-scores characterize the channel;
Cohen's \(d\) and Bayesian sample counts characterize
the adversary's decision problem under a 255/256 prior for
full-length keys.}%
\label{tab:keydist_stats}
\phantom{z}\\
\centering
\resizebox{1.0\linewidth}{!}{%
\begin{tabular}{lrrr}
\hline%
\textbf{Metric} & \textbf{OpenSSL} & \textbf{BoringSSL} & \textbf{BoringSSL} \\
       & \textbf{(stock)} & \textbf{(stock)}   & \textbf{(patched)}  \\
\hline%
MI, 33-class (bits) & 0.0031 & 0.6104 & 0.0055 \\
MI permutation \(z\)-score & 249 & 57{,}861 & 434 \\
Cohen's \(d\), binary & 0.098 & 1.11 & 0.008 \\
Binary MI (bits) & 0.0003 & 0.037 & 0.0006 \\
Binary MI \(z\)-score & 151 & 19{,}021 & 250 \\
Samples to \(P > 0.5\) & \(\approx\)1{,}000 & \(\approx\)10 & \(\approx\)100{,}000 \\
Samples to \(P > 0.999\) & \(\approx\)10{,}000 & \(\approx\)50 & \(\approx\)1{,}000{,}000 \\
\hline%
\end{tabular}
}
\end{table}

The 33-class MI characterizes the channel,
but the adversary's task is rather to decide if a key is
full length or not.
Partitioning into 384-bit keys versus all shorter encodings,
Cohen's \(d\) (where values near zero indicate overlap and values above one indicate separation)
is 0.098 for OpenSSL,
meaning the distributions are nearly indistinguishable,
and 1.11 for BoringSSL,
meaning the distributions barely overlap.
This interpretation is consistent with the binary MI permutation \(z\)-score
(standard deviations above the null hypothesis),
which are 151 for OpenSSL and 19,021 for BoringSSL\null.
All reject the null hypothesis,
but by this metric the BoringSSL leak is two orders of magnitude stronger
than that of OpenSSL\null.

For a randomly generated P-384 key,
the probability of the scalar having full byte length is 255/256,
a somewhat low probability for observing a truncated key.
Nevertheless, this true a priori probability reflects the adversary's actual decision problem.
For OpenSSL, the adversary requires approximately 1,000
observations to reach a posterior probability \(P>0.5\),
and 10,000 for \(P>0.999\).
For BoringSSL, averaging roughly 10 observations yields \(P>0.5\),
and 50 for \(P>0.999\),
two orders of magnitude less effort than OpenSSL\null.

\subsection{Disclosure, Defense, and Discussion}
We shared our findings with the BoringSSL security
team\footurl{https://marc.info/?l=oss-security&m=176046647932425}.
The remainder of this section discusses their response,
a simple hedge that we prototyped,
and the corresponding empirical results.

\PARAGRAPH{What BoringSSL could've done}
The leakage in BoringSSL spans the entire decode stack,
where PEM base64 decoding, DER parsing, and BIGNUM conversion
all process variable-length input proportional to the encoded key size.
Achieving constant-time behavior across these layers would require
invasive changes to the serialization infrastructure,
which the implementation currently does not pursue.

As an alternative, we implement a timing hedge internal
to \texttt{PEM\_read\_bio\_PrivateKey}.
On each call,
it spins up a second thread and runs the real parse concurrently with that of a fixed,
full-length private key,
and returns only after both complete (post-join).
The observed time is therefore dominated by the slower parse,
which (in practice) will approach the time of a full-length parse.
We acknowledge this strategy violates most of the models in \autoref{tab:compared},
yet the goal of this experiment is to raise the cost for adversaries,
not to eliminate all potential leakage.

We then gathered a third dataset from the methodology
in \autoref{sec:keyload_results} against this patched BoringSSL version.
The hedged BoringSSL distributions (\autoref{fig:keydist_boringssl}, right)
overlap, visually similar to OpenSSL\null.
Regarding information-theoretic metrics (\autoref{tab:keydist_stats}),
33-class MI drops from 0.61 to 0.005 bits
(a 110\(\times\) reduction, versus 0.003 for OpenSSL),
and binary MI from 0.037 to 0.0006 (versus 0.0003 for OpenSSL).
On the adversary's decision problem (short versus full) the hedge improves further.
Cohen's \(d\) drops from 1.11 to 0.008, below OpenSSL's 0.098,
and the adversary now requires roughly 100,000 observations to reach
\(P>0.5\) and 1,000,000 for \(P>0.999\),
notably four orders of magnitude more effort than against unpatched BoringSSL
and two orders of magnitude more than even OpenSSL\null.
Permutation testing for the null hypothesis still detects a signal across all datasets,
yet our patched BoringSSL is harder to target than OpenSSL
across every Bayesian metric.

\PARAGRAPH{What BoringSSL did}
During disclosure,
BoringSSL acknowledged the
leak but declined to react with any code changes.
On one hand, rejecting these malformed keys has unknown consequences
since the key landscape is not well understood,
which is a valid concern from an ecosystem and business perspective.
Compounding this, in their view non-conformant keys fall outside
their constant time boundary,
which they subsequently clarified means constant time only for fixed-width inputs
(see the bottom row of \autoref{tab:compared}).
The leak therefore falls outside their model by construction,
and for variable-width inputs it follows that any mitigation that continues to accept malformed
keys would also fail to satisfy the traditional constant time model.
On the other hand, our data in \autoref{tab:keydist_stats} offers an
adversary-driven perspective on the vulnerability.
Even our simple hedge that clearly still violates the traditional constant time model
increases the adversary's cost by four orders of magnitude,
and it does so while targeting only the OpenSSL model,
the weakest of all in \autoref{tab:compared}.
Counterintuitively, BoringSSL aims at considerably higher security in their model,
yet the leak is measurably stronger than in OpenSSL,
which applies a significantly weaker model.
In 1772, Voltaire wrote%
\footnote{Voltaire, ``La B\'egueule'' (1772)}
``le mieux est l'ennemi du bien'' or \emph{the best is the enemy of the good}.
Four orders of magnitude is a quantitative, objective improvement,
and we claim that practical hedges have a useful place in security hardening,
over arguments about model semantics.
\section{Conclusion}\label{sec:conclusion}

This work systematizes the evolution of cryptographic constant time models
in security-critical libraries,
from early embedded-era assumptions through modern project-specific policies.
The historical pattern is consistent,
showing every relaxation of the traditional model has eventually been exploited.
The traditional model's strength is its simplicity,
where all addresses are public,
making fewer assumptions about hardware behavior than any of the alternatives.

The gap between what constant time models protect and what specifications
assume is a recurring source of vulnerabilities.
Padding oracles, encoding leakage, and canonicalization issues all
originate in specifications that mandate operations on secrets that are
difficult or impossible to realize in constant time.
The methodology we distill from these case studies provides a
systematic framework for exploring this gap.

Applying the methodology to EC private key loading illustrates the
limits of threat model strength as a proxy for implementation security.
BoringSSL's constant time model is explicitly stronger than OpenSSL's,
in that it aims to protect against local and microarchitectural attackers,
while OpenSSL's security policy excludes colocation attacks entirely.
Yet on this specific leak,
our data from the adversarial perspective demonstrates that
OpenSSL is empirically more resistant.

We close with the following joke posted to rec.humor.funny
by Wolfgang Sohrt in 1989.

\begin{quote}
An engineer, a physicist and a mathematician have to build a fence around
a flock of sheep, using as little material as possible.

The engineer forms the flock into a circular shape and constructs a fence around it.

The physicist builds a fence with an infinite diameter
and pulls it together until it fits around the flock.

The mathematician thinks for a while,
then builds a fence around himself and defines himself as being outside.
\end{quote}

A strong threat model on paper means little if the
boundaries are drawn too narrowly.
Specifications exist in the wild,
non-conformant inputs exist in the wild,
and attackers do not respect scope declarations.
Empirical measurement, not model coverage, is the final arbiter
of whether an implementation is vulnerable to timing attacks.

\PARAGRAPH{Availability}
In support of Open Science and to ensure reproducibility,
we released\footurl{https://gitlab.com/platsec/boringssl-keyload-vuln}
our tooling from the data procurement methodology in \autoref{sec:keyload_results}
as a freely available research artifact for the benefit of the community.
The artifact can be used to reproduce at least
\autoref{fig:keydist_boringssl} (left, center).

\bibliographystyle{plainnat}%

\end{document}